\documentclass[structabstract]{aa} 
\usepackage{subfigure}                                   
\usepackage{epsfig}
\usepackage{natbib, graphicx}
\usepackage{amssymb, amsmath}
\usepackage{lscape}
\usepackage{multirow}

\usepackage{amsmath}
\usepackage{color}
\providecommand{\tabularnewline}{\\}
\usepackage{graphicx}
\usepackage{txfonts}
\begin{document}

\title{Kinematics and chemical properties of the \\
Galactic stellar populations\thanks{Based on observations collected at the La Silla Paranal Observatory,
ESO (Chile) with the HARPS spectrograph at the 3.6-m telescope (ESO
runs ID 72.C-0488, 082.C-0212, and 085.C-0063).}}

\subtitle{The HARPS FGK dwarfs sample}

\titlerunning{Kinematics and chemical properties of the Galactic disks}


\author{V.~Zh.~Adibekyan\inst{1} 
\and P.~ Figueira\inst{1}
\and N.~C.~Santos\inst{1,2}
\and A.~A.~Hakobyan\inst{3,4}
\and S.~G.~Sousa\inst{1,5}
\and G.~Pace\inst{1}
\and \\ E.~Delgado~Mena\inst{1}
\and A.~C.~Robin\inst{6}
\and G.~Israelian\inst{5,7}
\and J.~I.~Gonz\'{a}lez Hern\'{a}ndez \inst{5,7}
}

\institute{Centro de Astrof\'{\i}ísica da Universidade do Porto, Rua das Estrelas,
4150-762 Porto, Portugal\\
\email{Vardan.Adibekyan@astro.up.pt}
\and Departamento de F\'{\i}ísica e Astronomia, Faculdade de Ci\^{e}ncias da Universidade do Porto, Portugal
\and Byurakan Astrophysical Observatory, 0213 Byurakan, Aragatsotn province, Armenia
\and Isaac Newton Institute of Chile, Armenian Branch, 0213 Byurakan, Aragatsotn province, Armenia
\and Instituto de Astrof\'{\i}sica de Canarias, 38200 La Laguna, Tenerife, Spain
\and Institut Utinam, CNRS UMR6213, Universit\'{e} de Franche-Comt\'{e}, OSU THETA de Franche-Comt\'e-Bourgognen, Besan\c{c}on, France
\and Departamento de Astrof{\'\i}sica, Universidad de La Laguna, 38206 La Laguna, Tenerife, Spain}

   \date{Received ... / Accepted ...}

 
  \abstract
   {}
   {We analyze chemical and kinematical properties of about 850 FGK solar neighborhood  long-lived dwarfs observed with the HARPS high-resolution
 spectrograph. The stars in the sample have $\log\,g$ $\geq$ 4 dex, 5000 $\leq$ \emph{$T{}_{\mathrm{eff}}$} $\leq$ 6500 K, and -1.39 $\leq$ [Fe/H] $\leq$ 0.55 dex.
The aim of this study is to characterize and explore the kinematics and chemical properties of stellar populations of the Galaxy in order to understand 
their origins and evolution. }
   {We apply a purely chemical analysis approach based on the [$\alpha$/Fe] vs. [Fe/H] plot to separate Galactic stellar populations into the thin
disk, thick disk and high-$\alpha$ metal-rich  (h$\alpha$mr). Then, we explore the population's stellar orbital eccentricity distributions, their correlation with metallicity, and 
rotational velocity gradients with metallicity in the Galactic disks to provide constraints on the various formation models.}
   {We identified a gap in the [$\alpha$/Fe] - [Fe/H] plane for the $\alpha$-enhanced stars, and by performing a bootstrapped Monte Carlo test   
we obtained a probability higher than 99.99\% that this gap is not due to small-number statistics.
Our analysis  shows a negative gradient of the rotational velocity of the thin disk stars with [Fe/H] (-17 km~s$^{-1}$~dex$^{-1}$), and a steep positive gradient
 for both the thick disk and h$\alpha$mr stars with the same magnitude of about +42 km~s$^{-1}$~dex$^{-1}$. For the thin disk stars we observed no correlation between orbital 
eccentricities and metallicity, but observed a steep negative gradient for the thick disk and h$\alpha$mr stars with practically the same magnitude ($\approx$ -0.18 dex$^{-1}$).
 The correlations observed for the nearby stars (on average 45 pc)  using high-precision data in general agree well with the results obtained for the SDSS 
sample of stars located further from the Galactic plane.}
  {Our results suggest that radial migration played an important role in the formation and evolution of the thin disk. For the thick disk stars it is not possible 
to reach a firm conclusion about their origin. Based on the eccentricity distribution of the thick disk stars only their accretion origin can be ruled out, and 
the heating and migration scenario could explain the positive steep gradient of $V_{\phi}$ with [Fe/H]. 
{Analyzing the h$\alpha$mr stellar population we found that they
share properties of both the thin and thick disk population. A comparison of the properties of the h$\alpha$mr stars with that of the  
subsample of stars from the N-body/SPH simulation 
using radial migration suggest that they may have originated from the inner Galaxy.
Further detailed investigations would help to clarify their exact nature and origin.}}


\maketitle
%

\section{Introduction}

The  formation and  evolution of  the Galactic  disks is  an important
topic  in  contemporary  astrophysics.   The  Milky  Way  (MW)  has  a
composite  and complex  structure  with several  main components  (halo, bulge,  thin disk, and thick disk)
and  many stellar streams
with origin far from being thoroughly understood.

The main three stellar populations of the MW in the solar neighborhood
are the thin disk, the thick disk, and the halo, although most of these stars
belong to the thin and thick disks.  The subdivision  between the
thick disk  and thin disk was first  identified by \citet{Gilmore-83},
who  analyzed  the  stellar  density  distribution as  a  function  of
distance from  the Galactic  plane.  These two populations  have different
kinematics  and chemical  properties.   Generally, the  thick disk  is thought to be
composed   of  relatively  old stars \citep[e.g.][]{Bensby-03,  Bensby-05,
  Fuhrmann-08,   Adibekyan-11},   metal-poor,   and   $\alpha$-enhanced
\citep[e.g.][]{Fuhrmann-98,  Fuhrmann-08,  Prochaska-00,  Feltzing-03,
  Reddy-06,  Haywood-08,  Lee-11,  Adibekyan-12}   that  move  in
Galactic  orbits  with  a  large-scale height  and  long-scale  length
\citep[e.g.][]{Robin-96, Buser-99,  Juric-08, Kordopatis-11}. However,
some of  the observed  trends may depend  somewhat on the  criteria
applied   to   divide   stars    into   the   two   disk   populations
\citep[e.g.][]{Fuhrmann-08, Schonrich-09b, Lee-11}. Interestingly, this model has been recently challenged.  Recent analyses of
the  geometric  decompositions  of  the  Galactic disk  based  on  the
elemental-abundance selection  of the sample  stars yielded strikingly
different    results   \citep[see][]{Bovy-12a,    Bovy-12b,   Liu-12}.
\citet{Bovy-12a}  find  that  mass-weighted scale-height  distribution
smoothly varies  and there is  no thin-thick disk bi-modality  (i.e., the MW
has no distinct thick disk).  We refer the reader to \citet{Ivezic-12}
and \citet{Rix-13} for more recent  reviews of the stellar disk(s) and
populations.

While considering  that the thick disk  has formed only  by one single
mechanism  is  most   probably  too  semplicistic,  several  different
scenarios have been proposed for  the formation of thick disk:  heating of  a pre-existing  old  thin disk  via minor  mergers
\citep[e.g.][]{Quinn-93,Villalobos-08}, direct accretion of stars from
disrupted satellites  \citep[e.g.][]{Abadi-03}, gas accretion  at high
redshift and stars formed in situ \citep[e.g.][]{Brook-05}, and radial
migration    of     stellar    orbits    \citep[e.g.][]{Schonrich-09a,
  Schonrich-09b, Loebman-11}.

\begin{table*}
\centering
\caption{The average values of the $U{}_{\mathrm{LSR}}$, $V{}_{\mathrm{LSR}}$, and $W{}_{\mathrm{LSR}}$ velocity components and their standard deviations 
for the stellar groups, along with the number of stars and their percentage. The values obtained from the Gaussian fiting of the data presneted in brackets.}
\label{table-sigma}
\begin{tabular}{cccccccccc}
\hline
\hline
\noalign{\vskip0.01\columnwidth}
{Stellar groups} & {\footnotesize $U{}_{\mathrm{LSR}}$} & {\footnotesize  $\sigma{}_{\mathrm{U}}$} & {\footnotesize $V{}_{\mathrm{LSR}}$} 
& {\footnotesize  $\sigma{}_{\mathrm{V}}$} & {\footnotesize $W{}_{\mathrm{LSR}}$} & {\footnotesize $\sigma{}_{\mathrm{W}}$} & {\footnotesize [Fe/H]} & {\footnotesize $N$} & {\footnotesize Percentage}\tabularnewline
\hline
\noalign{\vskip0.01\columnwidth}
{{\footnotesize Thick}}  & {\footnotesize -5(-5)} & {\footnotesize 66(56)} & {\footnotesize -52(-53)} & {\footnotesize 39(42)} & {\footnotesize -8(-22)} & {\footnotesize 44(36)} & {\footnotesize -0.60$\pm$0.23(-0.58)} & {\footnotesize $84$} & {\footnotesize 9.94$\pm$1.03}\tabularnewline
{{\footnotesize h$\alpha$mr}}  & {\footnotesize -7(-3)} & {\footnotesize 42(39)} & {\footnotesize -23(-18)} & {\footnotesize 26(28)} & {\footnotesize 0(-2)} & {\footnotesize 21(20)} & {\footnotesize 0.03$\pm$0.12(0.03)} & {\footnotesize $58$} & {\footnotesize 6.85$\pm$0.86}\tabularnewline
{{\footnotesize Thick+h$\alpha$mr}}  & {\footnotesize -6(-5)} & {\footnotesize 58(45)} & {\footnotesize -40(-35)} & {\footnotesize 37(41)} & {\footnotesize -4(-10)} & {\footnotesize 36(32)} & {\footnotesize -0.34$\pm$0.37} & {\footnotesize $142$} & {\footnotesize 16.78$\pm$1.28}\tabularnewline
{{\footnotesize Thin}}  & {\footnotesize 0(-3)} & {\footnotesize 37(37)} & {\footnotesize -9(-6)} & {\footnotesize 22(23)} & {\footnotesize 0(-1)} & {\footnotesize 18(16)} & {\footnotesize -0.06$\pm$0.22(-0.05)} & {\footnotesize $692$} & {\footnotesize 81.79$\pm$1.32}\tabularnewline
{{\footnotesize Halo}}  & {\footnotesize -39} & {\footnotesize 154} & {\footnotesize -198} & {\footnotesize 60} & {\footnotesize -27} & {\footnotesize 73} & {\footnotesize -0.82$\pm$0.23} & {\footnotesize $12$} & {\footnotesize 1.42$\pm$0.40}\tabularnewline

\hline 
\end{tabular}
\end{table*}

The  proposed scenarios  offer  very different  observationally testable
signatures,  hence  certain well-defined  properties  of the  Galactic
disks should be able  to distinguish between these possible scenarios.
In  this  paper,  to  provide  constraints on  the  suggested  various
formation  models,   we  explore  the   stellar  orbital  eccentricity
distributions     \citep[e.g.][]{Sales-09,    DiMatteo-11}, a possible
correlation    between     the    eccentricities    and    metallicity
\citep[see][]{Lee-11},   and   rotational   velocity  gradients   with
metallicity in the thin and thick disks
\citep[e.g.][]{Lee-11,  Kordopatis-11, Navarro-11, Liu-12}.

For our  analyses we  used the stellar  sample of 1111  long-lived FGK
dwarf  stars from \citet{Adibekyan-12}.  To separate  and characterize
the   different  Galactic   stellar  subsystems,   we  focus   on  the
[$\alpha$/Fe] ratio (here ``$\alpha$'' refers to the average abundance
of Mg, Si, and Ti).

The  outline of  this paper  is as  follows: In  Sect.~\ref{sample} we
introduce the  sample and present  the chemical criteria of  the disks
deviation. Results  and discussion of comparisons of  our results with
the  predictions of  different contemporary  disk  formation scenarios
follow  in Sect.~\ref{res_disc}. Finally,  in Sect.~\ref{conclusions},
we draw  our main conclusions.

\section{The sample and disk definition}
\label{sample}

It is  becoming increasingly clear  that a dissection of  the Galactic
disks  based  only on  stellar  abundances  is  superior to  kinematic
separation     \citep[e.g.][]{Navarro-11,     Lee-11,    Adibekyan-11,
  Liu-12}. This is because chemistry is a relatively more stable property of a star than the spatial positions
and kinematics. In this  analysis, to  separate the  thin-  and thick disk
stellar components, we used the position of the stars in the [$\alpha$/Fe]
- [Fe/H] plane.

\subsection{Sample selection and stellar parameters}

Our initial sample comprises 1111 FGK dwarfs with  high-resolution spectra observed with the HARPS spectrograph \citep{Mayor-03} at the
ESO 3.6-m telescope (La Silla, Chile). {It is a combination of three HARPS subsamples: HARPS-1 \citep{Mayor-03}, HARPS-2 \citep{LoCurto-10}, and
HARPS-4 \citep{Santos-11}. 
HARPS-1 is composed of 451 stars, 376 of which have been selected from the CORALIE volume-limited sample%
\footnote{The limit distance for the stars in this sample depends on the stellar spectral type, i.e., a color-dependent distance cutoff was applied.%
} \citep{Udry-00} for being non-active and low-rotating and the rest are southern confirmed exoplanet host stars.
HARPS-2 sample was compiled with stars in the solar neighborhood out to 57.5 pc from the Sun \citep{LoCurto-10}  to complete a planet-search survey previously started with
the CORALIE spectrograph. Then active, high-rotating, known binaries and variable stars were excluded which lead the sample to 582 stars \citep{Sousa-11b}.
It is worth to note, that for the CORALIE sample a color cut-off (\textit{B-V} $<$ 1.2) was made.
HARPS-4 sample consists of 97 stars and was compiled from the catalog of \cite{Nordstrom-04}, selecting stars with \textit{b - y} $>$ 0.33 and with photometric [Fe/H] 
between -1.5 and -0.5 dex
\citep{Sousa-11a}. We note that there are four stars in common between HARPS-1 and HARPS-4, and 14 stars in common between the HARPS-2 and HARPS-4 subsamples.}

\begin{figure}
  \centering
    \includegraphics[angle=0,width=1\linewidth]{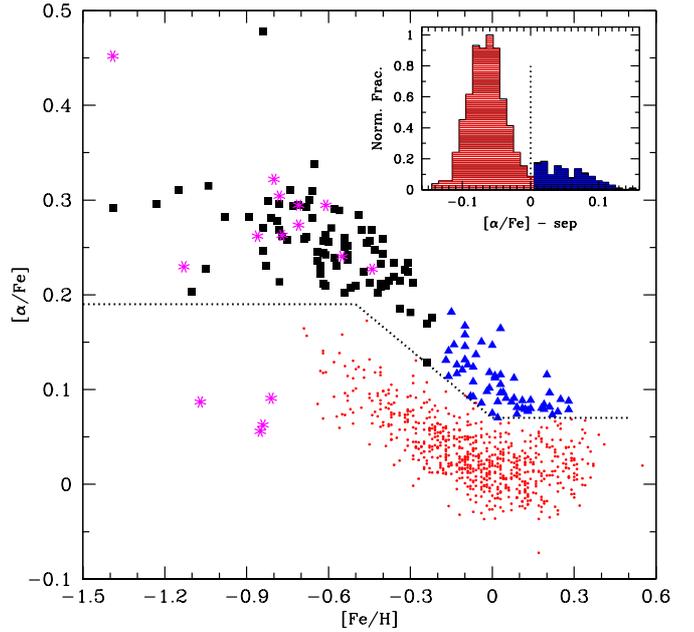}
    \caption{{[}$\alpha$/Fe{]} versus {[}Fe/H{]} for the whole sample. Magenta asterisks represent the stars belonging to the halo by their kinematics.
The black dotted line is the fiducial 
for division into thin- and thick disk populations. The black filled squares refer to the thick disk stars, blue triangles to the 
h$\alpha$mr, and the red dots to the thin disk stars. On the top-right corner the 
high-$\alpha$ and low-$\alpha$ separation histogram, after subtracting the separation curve, is plotted.}
\label{fig1}
\end{figure}

Precise stellar parameters (\emph{$T{}_{\mathrm{eff}}$}, $\log\,g$, [Fe/H], and 
\emph{$\xi{}_{\mathrm{t}}$}) and elemental abundances for 12 elements (Na, Mg, Al, Si, Ca, Ti, Cr, Ni, Co, Sc, Mn, and V) for all the stars 
were determined in an homogeneous manner.
Briefly summarizing, the equivalent widths of the lines were automatically measured with the ARES%
\footnote{The ARES code can be downloaded at http://www.astro.up.pt/sousasag/ares%
} code (Automatic Routine for line Equivalent widths in stellar Spectra - \citet{Sousa-07}). 
Then the atmospheric parameters and  elemental abundances were determined using a local thermodynamic equilibrium (LTE) 
analysis relative to the Sun with the 2010 revised version of the spectral synthesis code MOOG%
\footnote{The source code of MOOG can be downloaded at http://www.as.utexas.edu/$\sim$chris/moog.html%
} \citep{Sneden-73} and a grid of Kurucz ATLAS9 plane-parallel model atmospheres \citep{Kurucz-93}.
The reference abundances used in the abundance analysis were taken from \citet{Anders-89}. We refer the reader to \cite{Adibekyan-12} and \cite{Sousa-08}
for more details. 

In order to assemble a sample of stars with the most reliable and precise abundance determinations, we decided to establish a cutoff in temperature:
we excluded stars  with \emph{$T{}_{\mathrm{eff}}$} $\leq$ 5000 K and \emph{$T{}_{\mathrm{eff}}$} $\geq$ 6500 K. In those temperature regions
the errors, both in stellar parameters and [X/H] abundances, are higher \citep[see][for more details]{Neves-09, Sousa-11a, Sousa-11b, Adibekyan-12, Tsantaki-13}.
At the end of the establishing cutoff on temperature we finished with 869 stars. Since we are interested in long-lived dwarfs we excluded stars with  
$\log\,g$ $\leq$ 4 dex (23 stars), leading to a final sample of 846 stars.

Our final sample spans  the metallicity range -1.39 $\leq$ {[}Fe/H{]} $\leq$ 0.55 dex, although there are only nine stars with [Fe/H] $<$ -0.9 dex, and two stars
with [Fe/H] $>$ 0.4 dex. The typical relative uncertainties in the metallicity and [$\alpha$/Fe] are of about 0.03 dex \citep[see][]{Adibekyan-12}.

\begin{figure*}
\begin{center}$
\begin{tabular}{cc}
\includegraphics[angle=270,width=0.45\linewidth]{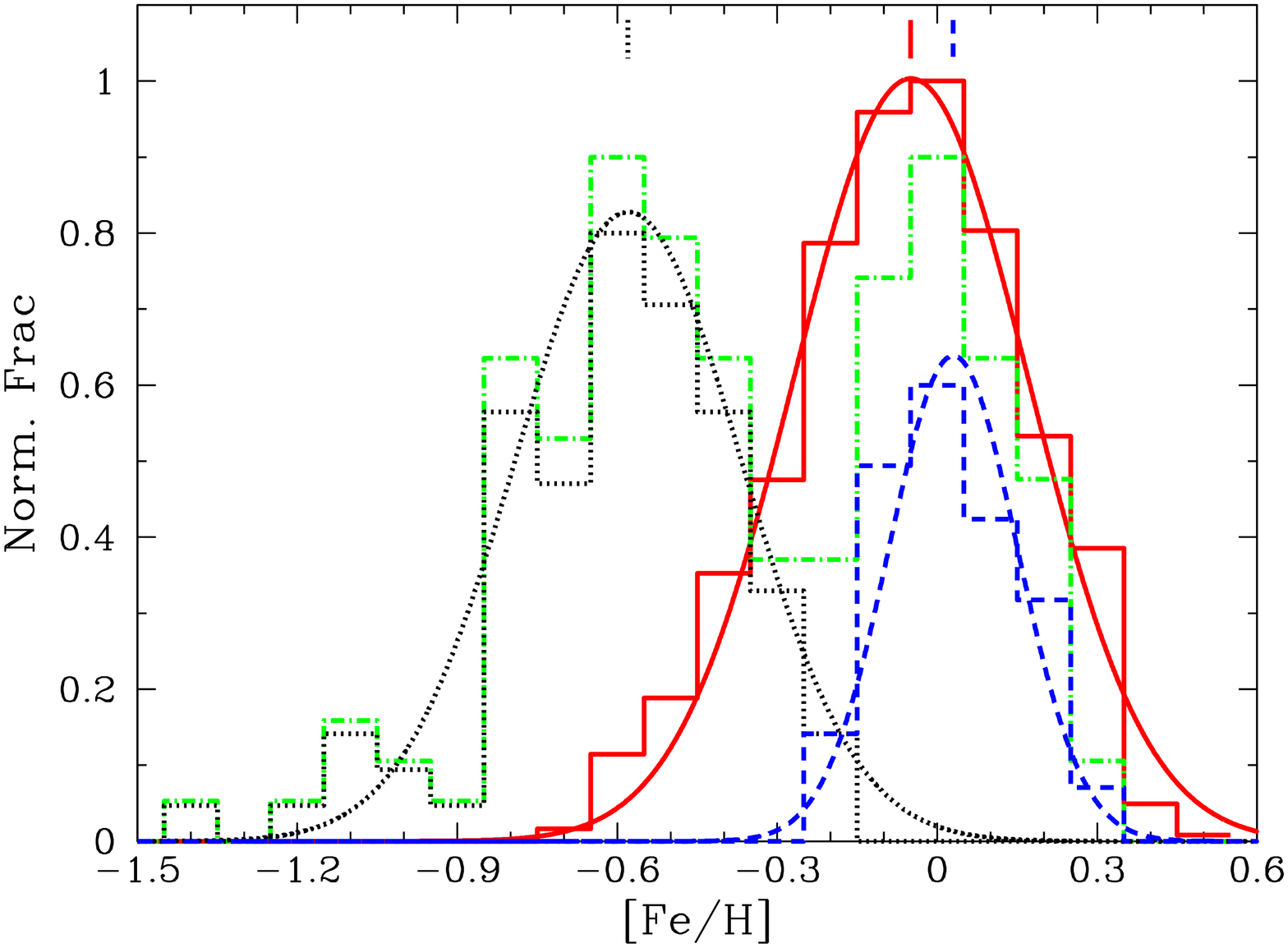}&
\includegraphics[angle=270,width=0.45\linewidth]{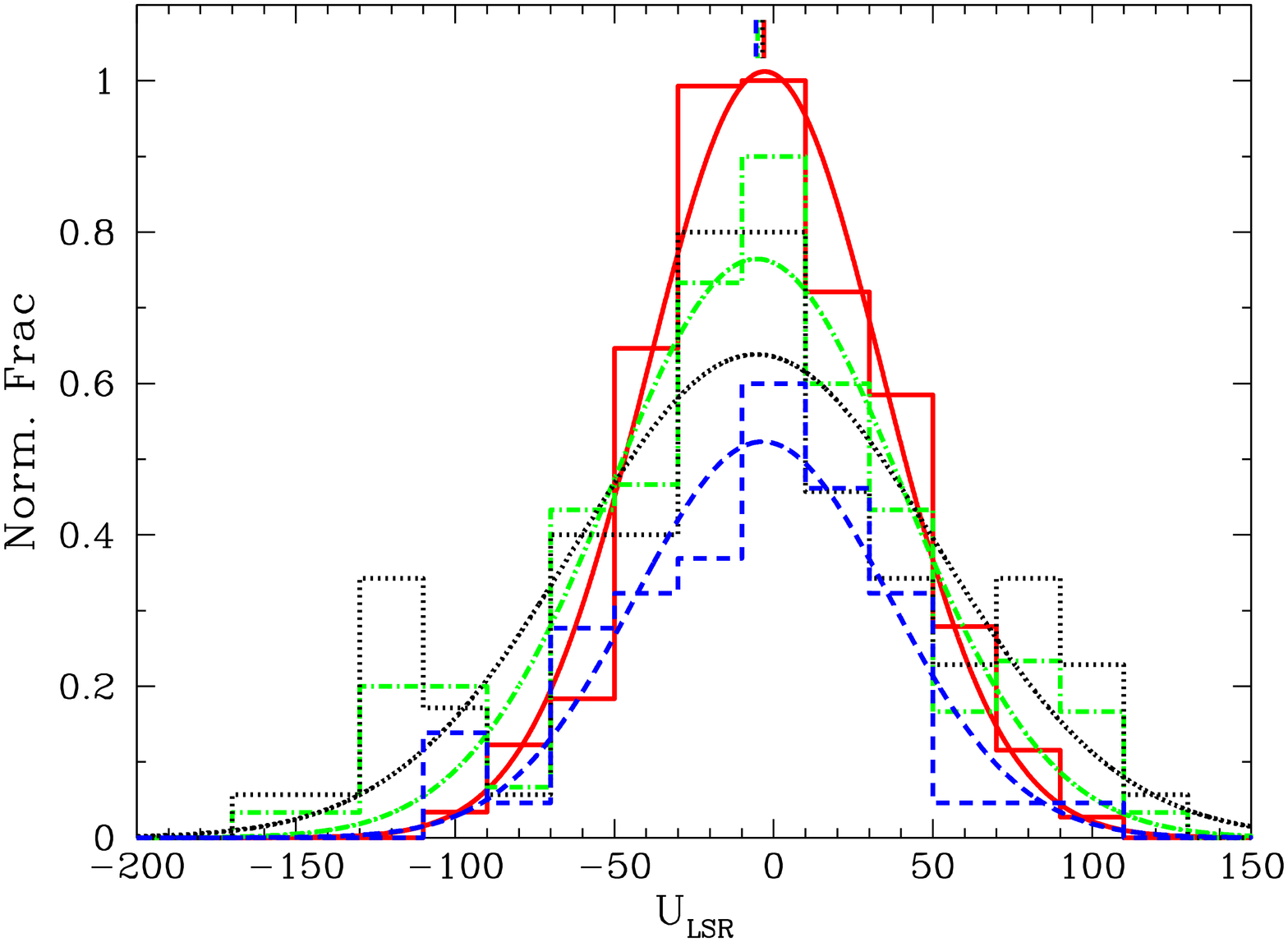}\tabularnewline
\includegraphics[angle=270,width=0.45\linewidth]{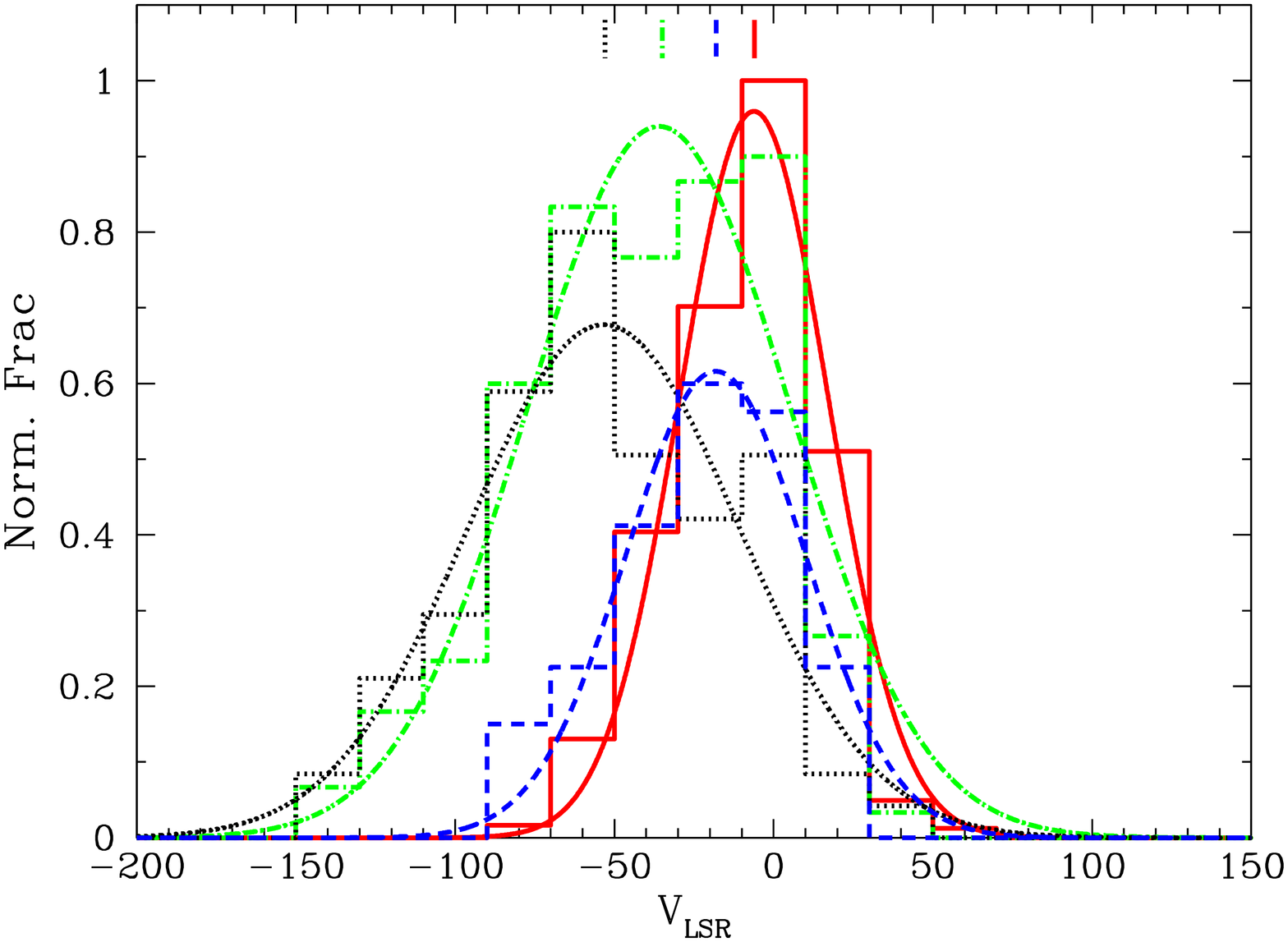}&
\includegraphics[angle=270,width=0.45\linewidth]{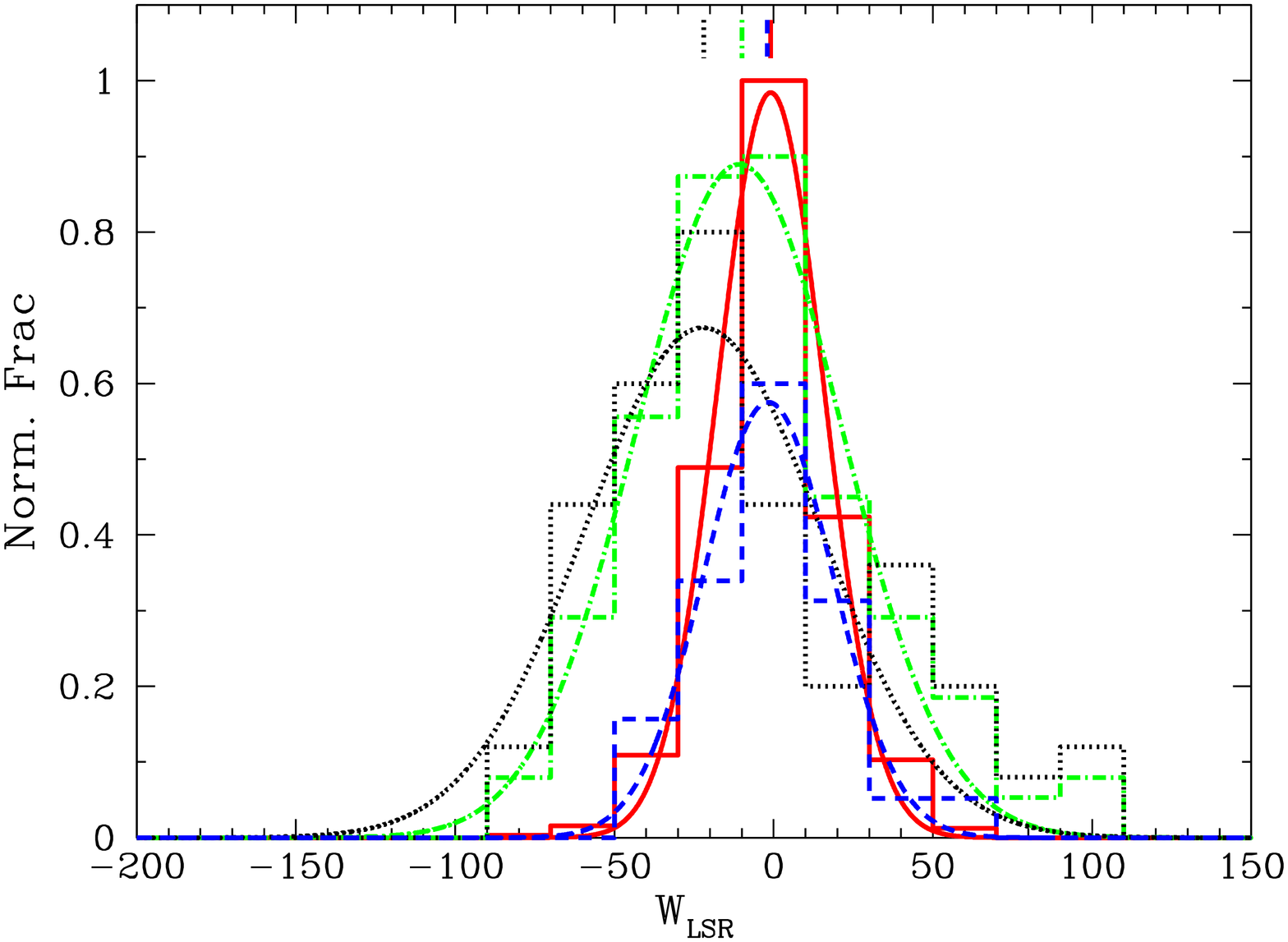}
\end{tabular}$
\end{center}
\caption{Distributions of the Galactic space velocities and metallicity of the stars from the thick disk (black dotted), h$\alpha$mr (blue dashed), 
the thin disk (red solid), and the thick disk + h$\alpha$mr (green dotted-dashed). Gaussian fits of the data are also presented.
The mean  of gaussian fits is also pictured by vertical lines on the top of each distribution. Note that the distributions of different groups were 
set lower/higher for the sake of clarity. }
\label{fig2}
\end{figure*}

\subsection{Dissecting the Galactic disks}

As mentioned above, the thin- and thick disk stars are different in their $\alpha$ content at a given metallicity ([Fe/H]). Here we use this dichotomy in 
the chemical evolution to separate different stellar populations.
 
\citet{Adibekyan-11} showed that the stars of this sample fall into two populations, clearly separated in terms of {[}$\alpha$/Fe{]} up to super-solar metallicities.
In turn, high-$\alpha$ stars were also separated into two families with a gap in both {[}$\alpha$/Fe{]} and metallicity ({[}$\alpha$/Fe{]}
$\approx$ 0.17 dex and {[}Fe/H{]} $\approx$ -0.2 dex) distributions. Here we use the same technique as in \citet{Adibekyan-11} to separate 
the stellar groups\footnote{
We note that in the current analysis we established a more conservative temperature cutoff than in our previous work.}.%

The [$\alpha$/Fe] versus [Fe/H] separation plot for the sample stars is depicted in Fig.~\ref{fig1}. 
The histogram (right-top corner) shows that the  separation between the stellar  families (low-$\alpha$ and high-$\alpha$) is clear.
Many recent papers separate the thin and thick disk stars by the [$\alpha$/Fe] ``gap'' at $\approx$ 0.2 - 0.3 dex \citep[see][]{Navarro-11, Lee-11} and 
other papers also report the existence of the ``knee'' in [$\alpha$/Fe] trends for the thick disk stars (kinematically selected) when [Fe/H] 
reaches to $\approx$ -0.3 dex \citep[e.g.][]{Feltzing-03,Bensby-07}. A similar gap in [$\alpha$/Fe] at $\approx$ 0.2 dex and similar
downturn in [Fe/H] at $\approx$ -0.3 dex are seen in Fig.~\ref{fig1}, and as already noted there is also a gap between metal-poor and 
metal-rich $\alpha$-enhanced stars {(the gap in [Fe/H] can be seen in top-left panel of Fig.~\ref{fig2})}. We note that there is no [Fe/H] gap found for the thin disk stars.
The presence of a metal-rich, old thick disk is already noticed in the spectroscopic studies of \cite{Feltzing-08} and \cite{Casagrande-12}, 
and a metal-rich $\alpha$-rich population has also been recently confirmed by \cite{Gazzano-13}. 

In the [Fe/H] region from -0.28 to -0.18 dex there are only three stars with enhanced $\alpha$ abundances, while at lower and higher 
metallicities the number of $\alpha$-enhanced stars is higher (9 and 13, respectively). In order to analyze whether this lack of high-$\alpha$ stars 
in the mentioned region is statistically significant or is due to a small-number statistics, we performed a simple bootstrapped Monte Carlo test.
We randomly drew 119 stars, which is the number of high-$\alpha$ stars in the metallicity region  -0.7 $\leq$ [Fe/H] $\leq$ 0.3 dex, from our sample in the same 
metallicity region (independent on their $\alpha$-content). Then we counted the number of stars in the interested metallicity region: 
from -0.28 to -0.18 dex. We also varied this interval by 0.05 dex ([-0.33; -0.23] to [-0.23; -0.13]). The entire process was repeated 10$^5$ times.
This test indicates that such a small amount of stars in the -0.28 to -0.18 dex metallicity region (or in the above mentioned intervals) can be obtained by chance with 
less than 0.01\% probability for four (or less) stars, and 0.05\% probability for five (or less) stars. In our simulation we had only one case where 
the mentioned metallicity interval contained three or less stars (0.001\%). This Monte Carlo test indicates that it is very unlikely that the observed gap between 
metal-poor and metal-rich $\alpha$-enhanced stars is due to small-number statistics and therefore the separation is considered real.

{As mentioned before, our sample consists of three HARPS subsamples, one of which (HARPS-4) was built to include only metal-poor stars.
Having this in mind, one might suspect that the observed gap could be a result of over-sampling of a thick disk metallicity distribution. 
However this can not be the case, because our thick disk sample includes only 32 stars from HARPS-4, and almost all of them have metallicities below -0.5 dex 
(there is only one star with [Fe/H] $>$ -0.4 dex, and two stars with -0.5 $<$ [Fe/H] $<$ -0.4 dex).}

The magenta asterisks in Fig.~\ref{fig1} refer to stars belonging to halo selected by their kinematics \citep[see][]{Adibekyan-11}. 
As is clearly evident, halo stars are divided into two high-$\alpha$ and
low-$\alpha$ groups. This dichotomy confirms the results already found by \cite{Nissen-10}.

\begin{table*}
\centering
\caption{Observed gradients, correlation coefficients of $V_{\phi}$ and eccentricity with metallicity, the number of stars, and significance of the correlation for 
 different stellar populations.}
\label{table-gradients}
\begin{tabular}{ccccccccc}
\hline
\hline
\noalign{\vskip0.01\columnwidth}
{Stellar groups} & {\footnotesize $\partial V_{\phi}/\partial\textnormal{[Fe/H]}$} & {r$^{}_{\partial V_{\phi}/\partial\textnormal{[Fe/H]}}$} & N$(V_{\phi})$ & n$\sigma (V_{\phi})$ & {\footnotesize $\partial e/\partial\textnormal{[Fe/H]}$} & {r$^{}_{\partial e/\partial\textnormal{[Fe/H]}}$} & N($e$) & n$\sigma$($e$) \tabularnewline
{ } & {(km s$^{-1}$ dex$^{-1}$)} & { } & & & {(dex$^{-1}$)} & { } & & \tabularnewline
\hline
\noalign{\vskip0.01\columnwidth}
{{\footnotesize Thick}}  & {\footnotesize 41.9$\pm$18.1} & {\footnotesize 0.247} & 84 & 2.24$\sigma$ & {\footnotesize -0.184$\pm$0.078} & {\footnotesize -0.266} & 73  & 2.24$\sigma$ \tabularnewline
{{\footnotesize h$\alpha$mr}}  & {\footnotesize 41.4$\pm$28.2} & {\footnotesize 0.190} & 58 & 1.43$\sigma$ & {\footnotesize -0.185$\pm$0.138} & {\footnotesize -0.212} & 40 & 1.30$\sigma$ \tabularnewline
{{\footnotesize Thick+h$\alpha$mr}}  & {\footnotesize 43.9$\pm$7.6} & {\footnotesize 0.435} & 142 & 5.19$\sigma$ & {\footnotesize -0.208$\pm$0.036} & {\footnotesize -0.475} & 113 & 5.27$\sigma$ \tabularnewline
{{\footnotesize Thin}}  & {\footnotesize -16.8$\pm$3.7} & {\footnotesize -0.164} & 692 & 4.43$\sigma$ & {\footnotesize -0.023$\pm$0.015} & {\footnotesize -0.212} & 515 & 4.81$\sigma$ \tabularnewline

\hline 
\end{tabular}
\end{table*}

\subsection{Characterisation of the Galactic components}

Guided by the chemical separation, we can characterize the stellar families in terms of their kinematics and metallicity.
Fig.~\ref{fig2} shows the distributions of the $U_{\mathrm{LSR}}$, $V_{\mathrm{LSR}}$, and $W_{\mathrm{LSR}}$ space velocity components relative to the local standard
of rest (LSR) and iron abundance
for our stellar groups (for details of the computation of the space velocity components we refer the reader to \citet[][]{Adibekyan-12}). 
In the plot we also separately present the distributions of the thick disk + h$\alpha$mr sample stars because the link 
between these two stellar groups is still under debate (see Sect.~\ref{res_disc}). The average values of the [Fe/H] and $U{}_{\mathrm{LSR}}$, 
$V{}_{\mathrm{LSR}}$, and $W{}_{\mathrm{LSR}}$ velocity components  and their standard deviations, along with their fractions are presented in Table~1.
We also performed Gaussian fits of the distributions of $U{}_{\mathrm{LSR}}$, $V{}_{\mathrm{LSR}}$, $W{}_{\mathrm{LSR}}$ and [Fe/H] 
for each Galactic component, and present the mean values and dispersion in the same table (the values in brackets). We did not perform Gaussian fitting 
for the halo stars because of the small number of stars in this group. We also did not fit [Fe/H] distribution of thick disk + h$\alpha$mr family, 
because the distribution is obviously bimodal and cannot be fitted with one Gaussian.

From the Fig.~\ref{fig2} and the corresponding table it is obvious that the velocity distributions of the  h$\alpha$mr stars are intermediate (even more similar to 
those of the thin disk stars) between thin and thick disk population
stars  \citep[see also][]{Adibekyan-11}, but with their average velocities and standard deviations having intermediate values between the thin- and thick disk families.
The Kolmogorov-Smirnov (K-S) statistics predicts 0.52, 0.01, and 0.02  probabilities (P$_\textnormal{KS}$)  that h$\alpha$mr and thin disk stars come from the same
underlying distribution for
$U{}_{\mathrm{LSR}}$, $V{}_{\mathrm{LSR}}$, and $W{}_{\mathrm{LSR}}$. The same statistical test gives much lower probabilities when we compare 
the velocity distributions of the h$\alpha$mr and thick disk stars - $U{}_{\mathrm{LSR}}$ (P$_\textnormal{KS}$ $\approx$ 0.38),  $V{}_{\mathrm{LSR}}$ (P$_\textnormal{KS}$ 
$\approx$ 3$\times$10$^{-5}$), and $W{}_{\mathrm{LSR}}$ (P$_\textnormal{KS}$ $\approx$ 0.005).
From the plot it can be noted that the velocity distributions of the thick disk stars are not perfectly Gaussian. This can be seen also from the Table 1
where the arithmetic averages (and $rms$) and means of the Gaussian fits (and standard deviations) are different.
The kinematic parameters obtained for the thin and thick disks, in general, agree well with those obtained for kinematically selected thin and thick disks 
in the solar neighborhood \citep[e.g.][]{Bensby-03, Soubiran-03, Robin-03}, but  local normalization for the thick disk is a bit higher
\citep[but see, also][]{Soubiran-03,Mishenina-04, Kordopatis-11}. 

\begin{figure}
  \centering
    \includegraphics[angle=270,width=1\linewidth]{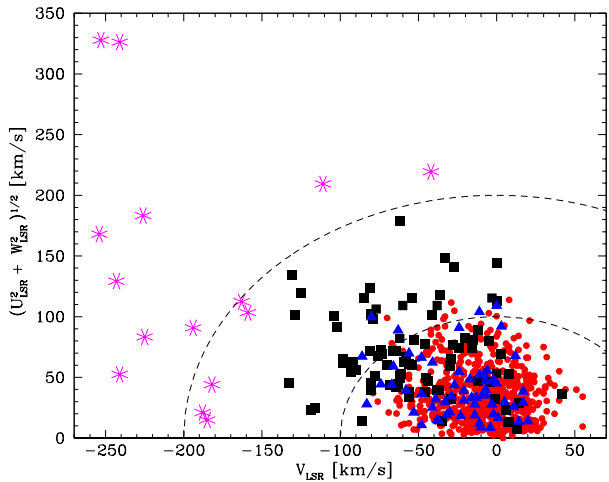}
    \caption{The Toomre diagram for the entire sample. The symbols are the same as in Fig. 1.}
\label{fig2.1}
\end{figure}

For illustrative purposes as a complement to Fig.~\ref{fig2}, the distribution of stars of our sample in the Toomre diagram is shown in Fig.~\ref{fig2.1}.

Although in Table 1 we present the average values of the velocities and metallicity obtained for halo stars, we note that these values should be 
considered with caution because the number of stars in the sample is very low. 

\section{Results and discussion}
\label{res_disc}

As mentioned in the Introduction, several scenarios have been proposed for the formation of the Galactic disks which should have left different
dynamical and chemical imprints. In this section we test the predictions of the mentioned scenarios using our sample of local thin- and thick disk 
FGK dwarfs as identified above. At the end of the section we discuss main properties of h$\alpha$mr stellar family.

\subsection{Rotational velocity and metallicity}
\label{rv_feh}

The large dispersion in the metallicities of the thin disk stars in the solar neighborhood, according to the migration-based scenarios \citep[]{Sellwood-02, Roskar-08,
Schonrich-09a, Minchev-10}, can be explained by inward radial migration of metal-poor and outward migration of metal-rich stars. A consequence of the radial migration
is a negative correlation between rotational velocities ($V_{\phi}$) and metallicity \citep[]{Schonrich-09a, Loebman-11} for the thin disk (low-[$\alpha$/Fe]) stars.
{We note that this prediction is not an ``exclusive'' to radial migration, models which assumed inside-out formation can also produce this anti-correlation.
This trend in the young stars arise from epicyclic motions of stars with their birth radii imprinted into their metallicities and disappears with time due to complete mixing.}
Hence, the same models predict small (positive) or no correlation between $V_{\phi}$ and [Fe/H] for stars with high-[$\alpha$/Fe] i.e. thick disk, old stars \citep[]{Loebman-11}.
Most of the recent observational studies using low-resolution spectra of large samples of thin disk stars outside the solar neighbourhood (mainly farther than 0.5 kpc from 
the Galactic disk) show negative gradient of $V_{\phi}$ with [Fe/H], which ranges from about -20 to -40  km s$^{-1}$ dex$^{-1}$ 
\citep[e.g.][]{Lee-11, Liu-12, Kordopatis-11}. The magnitude of the gradient also depends on the distance from the Galactic plane \citep[e.g.][]{Lee-11}. Interestingly, 
a recent study by \cite{Navarro-11} found little/no correlation between $V_{\phi}$  and metallicity for their 
thin disk stars in the solar neighborhood  based on a heterogeneous compilation of data from many different sources (their sample included also kinematically selected stars). 

\begin{figure}
\centering
\includegraphics[angle=270,width=1\linewidth]{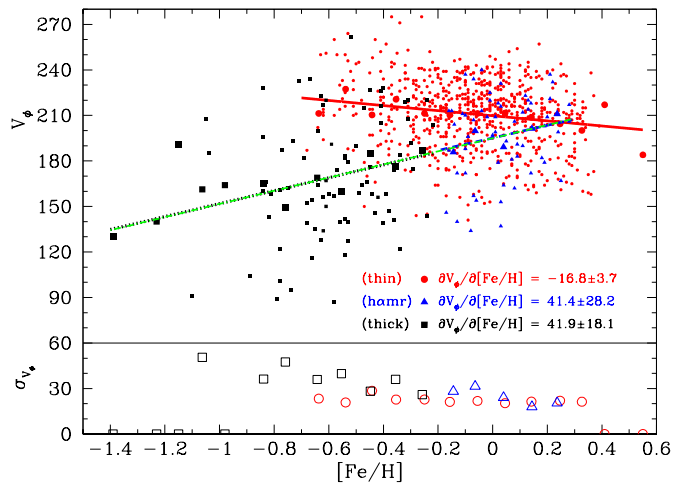}
\caption{Galactic rotational velocity (filled symbols) and velocity dispersion (open symbols) gradients with metallicity for stars assigned to different 
stellar populations. The symbols are the same as in previous plots. The smaller symbols correspond to the real stars and the larger symbols represent 
the $V_\phi$ and $\sigma{}_{\mathrm{V}}$ values for metallicity bins. Note that the slopes are obtained for the full unbinned data.}
\label{fig3}
\end{figure}

To study the degree of radial mixing by exploring the $V_{\phi}$-[Fe/H] correlation it is important to use a narrow range of galactocentric radius. Otherwise this correlation
will appear naturally due to the $V_{\phi}$  and metallicity gradients in the Galaxy. Our sample includes stars very close to the Sun (on average $\approx$ 45 pc) 
and is ideal for this search. 
The top panel of Fig.~\ref{fig3} shows the gradient of mean rotation velocity%
\footnote{We  adopt 8 kpc for the Sun's distance to the  Galactic center and 220 km s$^{-1}$ for the circular velocity of the LSR.%
} with metallicity for different groups of stars. The magnitudes of the gradients (the slopes) and 
the correlation coefficients are presented in Table 2.

The velocity gradient with metallicity obtained for the thin disk stars (-17 km s$^{-1}$ dex$^{-1}$)
is not far from the results (-22 km s$^{-1}$ dex$^{-1}$) obtained by \cite{Lee-11} for their thin disk stars in the range 0.1 $< \mid Z \mid <$ 0.5 kpc. 
We remind that this steep gradient is in disagreement with the result obtained by \cite{Navarro-11} for solar neighborhood thin disk stars. 

Recent studies obtained contradictory results analyzing $V_{\phi}$ gradient with [Fe/H] for the thick disk stars. A steep +40 to +50 km s$^{-1}$ dex$^{-1}$
gradient obtained by \cite{Spagna-10} and  \cite{Lee-11} using the SDSS data was confirmed by \cite{Kordopatis-11}, who used independent VLT/FLAMES observations of 700 thick disk
stars. Other studies \citep[e.g.][]{Ivezic-08, Bond-10, Loebman-11} did not reveal significant correlation probably due to larger errors in the photometric chemical 
abundances \citep{Lee-11}. We note that all the mentioned studies are based on the samples of stars located far from the Galactic plane (usually $\mid Z \mid > $ 0.5 kpc).

Fig.~\ref{fig3} and Table 2 indicate a steep and very similar gradients bot for the thick disk and h$\alpha$mr stars (about +42 km s$^{-1}$ dex$^{-1}$).
These values agree well with those obtained in 
\cite{Spagna-10, Lee-11, Kordopatis-11} and, as mentioned before, disagree with the theoretical predictions with radial migration by \cite{Schonrich-09a} and
\cite{Loebman-11}. {Interestingly, recent model of chemical enrichment in
the Galaxy by \cite{Curir-12} assuming an ``inverse'' gradient at high redshifts (z $>$ 3-4) explains the positive  rotation-metallicity correlation of the old thick disk 
population. They showed that using the inside-out formation and chemical evolution model of the Galactic disk suggested by \cite{Matteucci-89} and \cite{Chiappini-01},
this correlation can be established as a result of radial migration and heating processes of stars from the inner region of the
disk.}

\begin{figure}
  \centering
    \includegraphics[angle=270,width=1\linewidth]{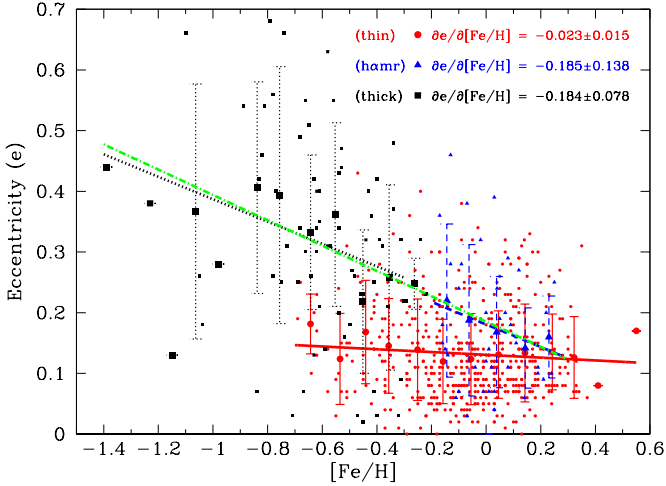}
    \caption{Trends of eccentricities as a function of metallicity for stars assigned to different stellar populations. 
The symbols are the same as in previous plots. The smaller symbols correspond to the real stars and the larger symbols present 
the eccentricities and standard deviations for each metallicity bin. The slopes are again obtained for the unbinned data.}
\label{fig4}
\end{figure}

\subsection{Stellar orbital eccentricities}
\label{e_feh}

The shape of eccentricity ($e$) distribution of thick disk stars can be used to constrain the main physical mechanism responsible for thier formation 
\citep[][]{Sales-09}. Recent studies \citep[e.g.][]{Dierickx-10, Wilson-11, Lee-11, Kordopatis-11}, which used the \cite{Sales-09} eccentricity test, 
reached a general conclusion that the eccentricity distribution is inconsistent with the accretion scenario, but could not distinguish between other published models, 
i.e. heating, migration or merger scenarios.

To discuss the distribution of the stellar orbital eccentricities we cross-matched our sample with the Geneva-Copenhagen Survey sample \citep{Casagrande-11},
which provides the orbital parameters of about 650 stars.

Fig.~\ref{fig4} shows trends of $e$ with [Fe/H] for different stellar families and the Table 2 presents the observed gradients quantitatively.
From Fig.~\ref{fig4} and corresponding table it is obvious that the thick disk and  h$\alpha$mr stars show very similar trends with the same slopes. On the other 
hand, the trend of the eccentricities for the thin disk stars is practically independent on metallicity. 
Our results quantitatively agree well with those obtained for the G-type dwarfs from the SDSS/SEGUE survey \citep{Lee-11}.

\begin{figure}
  \centering
    \includegraphics[angle=270,width=1\linewidth]{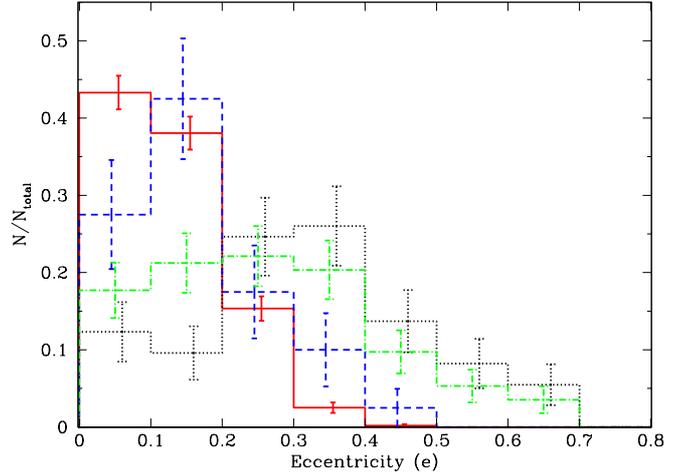}
    \caption{Normalized distributions of eccentricities for different stellar populations. Designation of the lines kept the same as in the previous plots.}
\label{fig5}
\end{figure}

Fig.~\ref{fig5} displays the normalized distributions of eccentricities for the different subsamples. The $e$ distribution of the thin disk stars peak at the first bin 
(the bin size is 0.1) and has an average value of 0.13. The distribution is also very concentrated around small eccentricities, with only 22 stars (4\%) having $e > 0.3$.
In contrast, the eccentricity distribution of the thick disk stars has a wider width and peaks at 0.3 $< e <$ 0.4 bin, with an average value of $e =$ 0.31. 
The eccentricities of the h$\alpha$mr stars have intermediate values between those of the thin- and thick disk stars, with an average value a bit closer to that of the 
thin disk ($e =$ 0.17). In case of h$\alpha$mr stars again very few stars (5 out of 40) have $e > 0.3$ and only one with $e > 0.4$. Joining the h$\alpha$mr and thick disk
subsamples we obtain an $e$ distribution with a peak at 0.2 $< e <$ 0.3 and average value of 0.26.

A comparison  with the model  predictions in \cite{Sales-09}  can only
rule out  the pure accretion scenario \citep{Abadi-03},  while it does
not strongly favour  or discard any of the  other three scenarios.  In
particular, the shape of the  $e$ distribution of the thick disk stars
does   not  differ   much  from   the  disk   heating  model   of  \cite{DiMatteo-11}.    
 
Although the h$\alpha$mr stars have smaller $e$ compared to the chemically defined thick disk stars, from the Fig.~\ref{fig5} it is difficult to conclude that they have
different origin or evolution. If the h$\alpha$mr stars just represent the metal-rich tail of the thick disk, then their smaller eccentricities are a natural
consequence of the negative gradient of $e$ with [Fe/H] observed for the thick disk stars (see Fig.~\ref{fig4}).

\subsection{Significance of the observed trends}
\label{significance}

To evaluate the statistical significance of the inferred correlations between $V_{\phi}$ vs. [Fe/H] and $e$ vs. [Fe/H] we used a bootstrap procedure. First, we obtain the zero-centered distribution of the correlation coefficient
by randomly bootstrapping (building random samples by shuffling the parameters among the observed set of parameters) the observed data pairs 
10$^4$ times. We calculate the correlation coefficient for each of these uncorrelated data sets and then
the average and standard deviation of these values. By assuming a Gaussian distribution for r$^{}_{\partial V_{\phi}/\partial\textnormal{[Fe/H]}}$ 
(r$^{}_{\partial e/\partial\textnormal{[Fe/H]}}$), we can
calculate the probability that the r$^{}_{\partial V_{\phi}/\partial\textnormal{[Fe/H]}}$ (r$^{}_{\partial e/\partial\textnormal{[Fe/H]}}$)
 of the original dataset was obtained by pure chance, but we merely present the offset in sigma (n$\sigma (V_{\phi})$ (n$\sigma (e)$) in Table 2).

From the table, one can note that the correlations between $V_{\phi}$ and [Fe/H] for h$\alpha$mr and thick disk stars separately are not very significant (1.43$\sigma$ and 2.24$\sigma$),
even if the slopes are very steep. Interestingly, the significance of the correlation increases a lot when we consider thick disk+h$\alpha$mr sample (5.19$\sigma$).
To check if this non-significance in both h$\alpha$mr and thick disk separate samples is due to small sizes of the samples or it is due to small metallicity ranges, we 
``merged'' the samples in the following way. Since the correlation coefficient (and its significance) does not change if one add or subtract a constant to the parameters 
(in this case to $V_{\phi}$ and [Fe/H]) we shifted the thick disk sample (i.e., shifted every point of it) to the \textit{locus} of h$\alpha$mr stars subtracting the difference 
between average values of [Fe/H] and $V_{\phi}$ of the thick disk and h$\alpha$mr stars%
\footnote{We remind the reader that the slopes of the correlation for the two samples are very similar.%
}. This procedure centers both data sets around the same point. Then we calculated the coefficient and significance of the correlation using the 
same bootstrapping approach.
For the  ``merged'' sample we obtained practically the same slope (41.8$\pm$14.6) as for the individual samples and correlation coefficient about 0.234. We found that this
correlation is significant at about 2.82$\sigma$ level. If we compare the later value with that obtained for the thick disk subsample we can note that by increasing the number 
of objects in the sample by about 70\% we did not increase significance of the correlation significantly. This suggests that the main reason of the obtained ``low'' 
significance is probably the small range of [Fe/H] for both independent thick disk and especially h$\alpha$mr subsamples. 

We used the same center-shifting approach (as we did for the $V_{\phi}$-[Fe/H] data pairs) to analyze the significance of the  correlation as a  function of the  
size of the  thick disk and h$\alpha$mr  subsamples. We found that increasing the size of the subsamples  (by about 55\% for the thick disk stars) does not significantly increase 
the significance of the correlations (from 2.24$\sigma$ to 2.71$\sigma$ for the thick disk subsample). This is probably once again due to small range of metallicities
of the samples.

{In our sample we have few thick disk stars with metallicities from -1.4 to -0.9 dex and these stars are widely spread in this metallicity range.
These rather extreme-value stars can influence the fitted slope significantly. 
From Figs.~\ref{fig3} and ~\ref{fig4} one can note that at lowest metallicities the stars (7 stars with [Fe/H] $<$ -0.9 dex)
mostly lie above the $V_{\phi}$-[Fe/H] correlation line and mostly below the $e$-[Fe/H] correlation line. Interestingly, the much larger data in  \cite{Lee-11} suggest similar behavior
for their most metal-poor thick disk stars \citep[see Figs. 7 and 9 of][]{Lee-11}. Nevertheless, to check how these data could alter the slopes we established a cutoff in
metallicity at [Fe/H] = -1.1, -1, and 0.9 dex. We found that the $V_{\phi}$-[Fe/H] slope changes from 48.7$\pm$21.9 to 61.5$\pm$25.8 km s$^{-1}$ dex$^{-1}$ and the slope of $e$-[Fe/H] varies 
from -0.26$\pm$0.09 to -0.35$\pm$0.11 dex$^{-1}$. This suggests that decreasing the lower metallicity limit for the thick disk the slopes become steeper staying within
the one-sigma errors.}

{From our tests we can conclude that the observed trends are real, but the exact values of the slopes (for the thick disk) should be considered with caution.}

\subsection{High-$\alpha$ metal-rich stars}
\label{significance}

{In this subsection we discuss some properties of h$\alpha$mr stellar family, which they share with thin- and thick disk population.}

From Fig.~\ref{fig3} it is interesting to see that although the h$\alpha$mr and the thin disk stars at [Fe/H] $>$ 0 dex have similar rotation
velocities, their trends with metallicity are completely different. The h$\alpha$mr stars behave like an extension of the thick disk stars on the $V_{\phi}$-[Fe/H] plane which
might mean a similar origin and/or evolution. The same similar behavior to the thick disk stars can be seen from From Fig.~\ref{fig4}. Although the 
h$\alpha$mr stars have orbits with eccentricities on average more similar to those of the thin disk stars (see Fig.~\ref{fig5}), the $e$ vs. [Fe/H] trend of 
these stars is the same as for the thick disk stars.

{Despite the mentioned similarities to the thick disk, the h$\alpha$mr stars also share some properties of the thin disk. From Table 1 one can see that the dispersion of
all the velocity components of h$\alpha$mr stars is very similar to those of the thin disk sample. In particular the low dispersion of the vertical velocities suggests a short
scale-height, very similar to that of the thin disk. \cite{Adibekyan-11} already showed that the average maximum vertical distance (Z$_{\rm{max}}$) the stars can reach above the 
Galactic plane is about 0.3 kpc for the h$\alpha$mr. For comparison the average values of Z$_{\rm{max}}$ are about 0.25 and 1 kpc for the thin disk and thick disk, respectively.
\cite{Adibekyan-11} also showed that the h$\alpha$mr and thick disk family stars have almost the same age, being on average older than thin disk stars by about 3 Gyr.
\cite{Adibekyan-12} analyzed the [X/Fe] trends with metallicity of  h$\alpha$mr stars for different elements and showed that for some elements the trends
are different than those of the thick disk stars.} 

At the same time, h$\alpha$mr stars are more metal-rich than the thick disk stars (if they do not just represent the metal-rich 
tail of the thick disk) and they are as metal-rich as the bulge stars and have similar enhanced $\alpha$-element abundances
compared to Galactic disk stars as recently found for the bulge stars \citep[e.g.][]{Fulbright-07, Bensby-13, Ness-13}%
\footnote{We note that the conclusion that  bulge stars are more enhanced in $\alpha$-elements is also based on the Fig. 27 of Bensby et al. 2013, but these authors did
not conclude that in their paper.%
}. 

{The mentioned results hint that h$\alpha$mr stars might have an origin in the inner Galaxy (bulge) and have migrated to the solar annulus \citep[see also][]{Adibekyan-11,
Gazzano-13}. Very recently, in their simulation \cite{Roskar-12} identified a sub-population with high-[Fe/H] and low-[O/Fe]. Although their simulations did not reproduce our data, i.e.
separation between low- and high-$\alpha$ stars at super-solar metallicities, at first glance the higher-[O/Fe] part of that family shares many properties of the h$\alpha$mr 
stars. From Fig. 7 of \cite{Roskar-12} and Fig. 14 of \cite{Loebman-11}%
\footnote{We note that in both papers they used the same N-body/SPH simulation.
} we can see that our h$\alpha$mr stars match well with the intermediate/old age population of stars with short scale-length and scale-height, migrated from the central part of
the Galaxy ($<$ 2 kpc).}

\section{Summary and conclusions}
\label{conclusions}

We selected a sample of about 850 solar neighborhood FGK dwarf stars with precise stellar velocity components and chemical abundances derived from the high-resolution
HARPS spectra. The stars have $\log\,g$ $\geq$ 4 dex, 5000 $\leq$ \emph{$T{}_{\mathrm{eff}}$} $\leq$ 6500 K, and -1.4 $<$ [Fe/H] $\leq$ 0.55 dex. 

Applying purely chemical approach
based on the [$\alpha$/Fe] vs. [Fe/H] plot, we separate Galactic stellar populations into the thin disk, thick disk, and h$\alpha$mr. The later two families are separated 
based on the observed ``gaps'' in both [Fe/H] and [$\alpha$/Fe]  distributions. Performed  bootstrapped Monte Carlo test gives a probability higher than 99.99\% that 
the observed ``gap'' is statistically significant and is not due to a small-number statistics. 

We characterize the stellar subsamples in terms of their kinematics and metallicity and found in general good agreement with earlier determinations of these parameters,
although the local normalization  for the thin disk is about 82\%, which is lower than the usually adopted value \citep[but see, also][]{Soubiran-03,Mishenina-04, Kordopatis-11}.

Our analysis  shows that the rotational velocity of the thin disk stars decreases with increasing [Fe/H] (-17 km s$^{-1}$ dex$^{-1}$), while this trend shows an 
increases with [Fe/H] for both thick disk and h$\alpha$mr, both with the same magnitude of about +42 km s$^{-1}$ dex$^{-1}$. {Although the magnitude 
of the slope for the thick disk stars depends on the cutoff established in [Fe/H].} These results, obtained for 
solar neighborhood stars (on average $\approx$ 45 pc), 
are an independent confirmation of those found for the SDSS sample of  stars located far from the Galactic plane \citep[e.g.][]{Lee-11}. The negative gradient of 
$V_{\phi}$ with [Fe/H] for the thin disk qualitatively agrees well with the expectations from the radial migration model and the steep positive gradients observed for 
the thick disk and h$\alpha$mr stars can be explained by the recent simulation by \cite{Curir-12} assuming an inverse chemical gradient in the inner early Galaxy.

For the thin disk stars we observed no correlation between orbital eccentricities and metallicity, but observed a steep negative gradient for the thick disk and 
h$\alpha$mr stars with practically the same magnitude of -0.185 dex$^{-1}$. {Again the magnitude 
of the slope for the thick disk stars depends on the cutoff established in [Fe/H].} These results quantitatively agree well with \cite{Lee-11}. The eccentricity distribution
of the thin disk subsample suggests that the radial migration played a dominant role in their formation and evolution. From the $e$ distribution of the thick disk stars
it is difficult to conclude which mechanism played the dominant role (heating, migration, or merger scenario), but the peak of the distribution observed at rather lower values 
than $e \sim$ 0.5 and the absence of the secondary peak at higher $e \sim$ 0.8 exclude the accretion origin. Probably
the combination of the two (or more) processes such as a heating and migration may describe better the observed distribution, just like it explains the positive rotational velocity gradient
with metallicity \citep{Curir-12}.

{As previously reported by \cite{Adibekyan-11} there is a family of stars with high [$\alpha$/Fe] values at solar and super-solar metallicities \citep[see also][]
{Gazzano-13}. Although the h$\alpha$mr stars have intermediate values (even closer to the thin disk) of the orbital parameters and space velocities between thin and 
thick disks, they  show the same behavior as the thick disk stars on the $V_{\phi}$ - [Fe/H] and $e$ - [Fe/H] planes.
These stars share properties of both the thin and thick disk populations. The properties of the h$\alpha$mr stars compares well with the metal-rich higher-[O/H] stars 
in the simulation by \cite{Loebman-11} and \cite{Roskar-12}, which have intermediate/old age and are extreme-migrators (probably $<$ 1-2 kpc).   
Based on the mentioned results we are inclined to think that the h$\alpha$mr stellar family may have originated from the inner Galactic disk/bulge and migrated 
up to solar neighborhood, although further investigations are needed to clarify their exact nature.}


%
\begin{acknowledgements}

{This work was supported by the European Research Council/European Community under the FP7 through Starting Grant agreement 
number 239953. V.Zh.A., S.G.S., and E.D.M are supported by grants SFRH/BPD/70574/2010, 
SFRH/BPD/47611/2008, and SFRH/BPD/76606/2011 from the FCT (Portugal), respectively.
G.P. is supported by grant SFRH/BPD/39254/2007 and by the project PTDC/CTE-AST/098528/2008, funded by FCT, Portugal.
G.I. and J.I.G.H. acknowledge financial support from the Spanish Ministry project MINECO AYA2011-29060, and J.I.G.H. also from the Spanish Ministry of Economy
and Competitiveness (MINECO) under the 2011 Severo Ochoa Program MINECO SEV-2011-0187.
{We gratefully acknowledge the Anonymous Referee for the constructive comments and suggestions.}}
\end{acknowledgements}

\bibliography{refbib}

\end{document}